\documentclass[prb,twocolumn,letterpaper]{revtex4-1}
\usepackage{bm,graphicx,graphics,amsmath,amssymb,bm,epsfig,color}
\usepackage{euscript,tabularx}
\usepackage{longtable}

\begin{document}

\title{Comparative Measurements of Inverse Spin Hall and
  Magnetoresistance \\ in YIG$|$Pt and YIG$|$Ta}

\author{C. Hahn}

\affiliation{Service de Physique de l'\'Etat Condens\'e (CNRS URA
  2464), CEA Saclay, 91191 Gif-sur-Yvette, France}

\email{christian.hahn@cea.fr}

\author{V.V. Naletov}

\affiliation{Service de Physique de l'\'Etat Condens\'e (CNRS URA
  2464), CEA Saclay, 91191 Gif-sur-Yvette, France}

\affiliation{Institute of Physics, Kazan Federal University, Kazan
  420008, Russian Federation}

\author{G. de Loubens}

\affiliation{Service de Physique de l'\'Etat Condens\'e (CNRS URA
  2464), CEA Saclay, 91191 Gif-sur-Yvette, France}

\email{gregoire.deloubens@cea.fr}

\author{O. Klein}

\affiliation{Service de Physique de l'\'Etat Condens\'e (CNRS URA
  2464), CEA Saclay, 91191 Gif-sur-Yvette, France}

\author{M. Viret}

\affiliation{Service de Physique de l'\'Etat Condens\'e (CNRS URA
  2464), CEA Saclay, 91191 Gif-sur-Yvette, France}

\author{J. Ben Youssef}

\affiliation{Universit\'e de Bretagne Occidentale, Laboratoire de
  Magn\'etisme de Bretagne CNRS, 6 Avenue Le Gorgeu, 29285 Brest,
  France}

\date{\today}

\begin{abstract}

  We report on a comparative study of spin Hall related effects and
  magnetoresistance in YIG$|$Pt and YIG$|$Ta bilayers. These combined
  measurements allow to estimate the characteristic transport
  parameters of both Pt and Ta layers juxtaposed to YIG: the spin
  mixing conductance $G_{\uparrow \downarrow}$ at the YIG$|$normal
  metal interface, the spin Hall angle $\Theta_{SH}$, and the spin
  diffusion length $\lambda_{sd}$ in the normal metal. The inverse
  spin Hall voltages generated in Pt and Ta by the pure spin current
  pumped from YIG excited at resonance confirm the opposite signs of
  spin Hall angles in these two materials. Moreover, from the
  dependence of the inverse spin Hall voltage on the Ta thickness, we
  extract the spin diffusion length in Ta, found to be
  $\lambda_{sd}^\text{Ta}=1.8\pm0.7$~nm. Both the YIG$|$Pt and
  YIG$|$Ta systems display a similar variation of resistance upon
  magnetic field orientation, which can be explained in the recently
  developed framework of spin Hall magnetoresistance.

\end{abstract}

\maketitle

\section{Introduction}

Spintronics aims at designing devices which capitalize on the
interplay between the spin- and charge-degrees of freedom of the
electron. In particular, it is of central interest to study the
interconversion from a spin current, the motion of spin angular
momentum, to a charge current and the transfer of spin angular
momentum between the conduction electrons of a normal metal (NM) and
the magnetization of a ferromagnetic material (FM). The separation of
oppositely spin polarized electrons of a charge-current through
spin-orbit-coupling is called spin Hall effect (SHE)
\cite{dyakonov71,hirsch99}. Its inverse process (ISHE) converts
spin-currents into charge-currents and has recently sparked an intense
research activity \cite{valenzuela06,kimura07}, as it allows for an
electrical detection of the dynamical state of a ferromagnet
\cite{saitoh06,kajiwara10}. Indeed, a precessing magnetization in a
ferromagnet generates a spin current via spin pumping
\cite{tserkovnyak05}, which can be converted, at the interface with an
adjacent normal layer, to a dc voltage by ISHE. Moreover, electronic
transport can also be affected by the static magnetization in the FM
as electrons spins separated by SHE can undergo different
spin-flip-scattering on the interface with the FM layer. In
particular, spin flipped electrons are deflected by ISHE in a
direction opposite to the initial current, leading to a reduced total
current at constant voltage. This effect depends on the relative
orientation between magnetization and current direction, and has
recently been called spin Hall magnetoresistance (SMR)
\cite{nakayama12}.

Experimental studies on spin pumping induced inverse spin Hall
voltages ($V_\text{ISH}$) in FM$|$NM bilayers were first carried out
with Pt as NM in combination with NiFe as FM
\cite{saitoh06,ando08,azevedo11,feng12,rousseau12} and more recently
with the insulating ferrimagnet Yttrium Iron Garnet (YIG)
\cite{kajiwara10,sandweg10,kurebayashi11,vilela-leao11,chumak12,castel12a}. Although
other strong spin-orbit metals have been tried in combination with the
metallic ferromagnets NiFe \cite{mosendz10a,kondou12} and CoFeB
\cite{liu12,pai12}, inverse spin Hall voltage
\cite{kajiwara10,castel12a} and magnetoresistance
\cite{nakayama12,huang12} measurements made on YIG$|$NM have so far
been limited to NM=Pt. Still, it would be very interesting to compare
$V_\text{ISH}$ and SMR measurements on different YIG$|$NM systems,
including metals having opposite spin Hall angles, such as Pt vs. Ta
\cite{morota11,liu12}. Ab initio calculations indeed predict the spin
Hall angle of the resistive $\beta$-phase of Ta to be larger and of
opposite sign to that of Pt \cite{tanaka08}. The defining parameters
for $V_\text{ISH}$ and SMR are the spin diffusion length in the normal
metal ($\lambda_{sd}$), the spin Hall angle ($\Theta_{SH}$) which
quantifies the efficiency of spin- to charge-current conversion, and
the spin mixing conductance ($G_{\uparrow \downarrow}$) which depends
on the scattering matrices for electrons at the FM$|$NM interface
\cite{tserkovnyak05} and can be seen as the transparency of the
interface for transfer of spin angular momentum \cite{jia11}.
Evaluation of the three above mentioned parameters is a delicate task
\cite{liu11b}, as the measured $V_\text{ISH}$ voltages and SMR ratio
depend on all of them.

In this paper, we present a comparative study of YIG$|$Pt and YIG$|$Ta
bilayers, where we measure both the ISHE and SMR on each sample.  We
confirm the opposite signs of spin Hall angles in Pt and Ta and the
origin of SMR, which has been explained in
Ref.\cite{nakayama12}. Thanks to these combined measurements, we can
evaluate the spin mixing conductances of the YIG$|$Pt and YIG$|$Ta
interfaces and the spin Hall angles in Pt and Ta. In order to get more
insight on the previously unexplored YIG$|$Ta system, we study the
dependence of ISHE on Ta film thickness, which enables us to extract
the spin diffusion length in Ta.

The remaining of the manuscript is organized as follows. Section II
gives details on the samples and experimental setup used in this
study. In section III, the experimental data of $V_\text{ISH}$ and SMR
obtained on the YIG$|$Pt and YIG$|$Ta systems are presented and
analyzed. In section IV, we discuss the transport parameters extracted
from our measurements. We also comment on the absence of direct effect
of a charge current in Pt on the linewidth of our 200~nm thick YIG
samples. Finally, we emphasize the main results of this work in the
conclusion.

\section{Experimental details}

\subsection{Samples}

\subsubsection{YIG films}

Two single crystal Y$_3$Fe$_5$O$_{12}$ (YIG) films of 200~nm thickness
were grown by liquid phase epitaxy on (111) Gd$_3$Ga$_5$O$_{12}$ (GGG)
substrates \cite{castel12b}, and labeled YIG1 and YIG2. Epitaxial
growth of the YIG was verified by X-ray diffraction and the films
roughness was determined by atomic force microscopy to be below
5~\AA. Their magnetic static properties were investigated by vibrating
sample magnetometry. The in-plane behavior of the thin YIG films is
isotropic with a coercitivity below 0.6~Oe \cite{castel12b}. The
saturation magnetization, found to be 140~emu/cm$^3$, corresponds to
the one of bulk YIG. This value was verified by performing
ferromagnetic resonance (FMR) at different excitation frequencies.

\begin{figure}
  \includegraphics[width=\columnwidth]{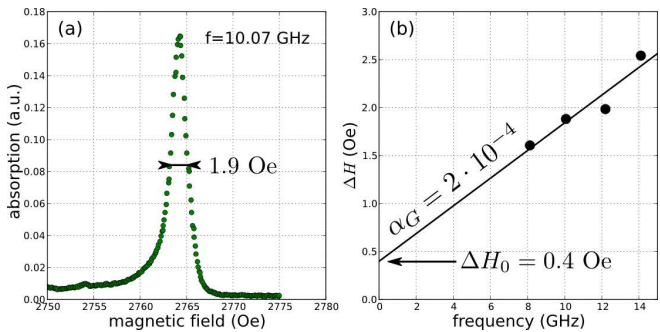}
  \caption{(Color online). (a) Standard in-plane FMR spectrum of a
    bare YIG 200~nm thin film used in this study. (b) Full FMR
    linewidth vs. frequency.}
  \label{fig:1}
\end{figure}

FMR also allows to extract the magnetic dynamic properties of the
200~nm thick YIG films. A typical FMR spectrum of the YIG1 film
obtained at 10~GHz and low microwave power ($P=-20$~dBm) is presented
in Fig.\ref{fig:1}a. The gyromagnetic ratio of our YIG films is found
to be $\gamma=1.79\cdot10^7$~rad/s/Oe. From the dependence of the
linewidth on the excitation frequency, their Gilbert damping $\alpha_G
= (2.0\pm0.2) \cdot 10^{-4}$ can be determined, see
Fig.\ref{fig:1}b. This value highlights the very small magnetic
relaxation of these thin films. Still, there is an inhomogeneous part
to the linewidth ($\Delta H_0=0.4$~Oe in Fig.\ref{fig:1}b). For one of
the two prepared films (YIG2), two to three closely spaced resonance
lines could be observed in some cases, which we attribute to distinct
sample regions having slightly different properties.

\subsubsection{YIG$|$Pt and YIG$|$Ta bilayers}

After these standard magnetic characterizations, the YIG films were
cut into slabs with lateral dimensions of 1.1~mm $\times$ 7~mm in
order to perform inverse spin Hall voltage and magnetoresistance
measurements. Platinum and tantalum thin films were then grown by
sputter deposition, at a power density of 4~W/cm$^2$. The growth of
the resistive $\beta$-phase Ta was achieved by optimizing the
Ar-pressure during the sputtering process. The appearance of this
tetragonal crystalline phase in a narrow window around $10^{-2}$~mbar
was verified by the presence of characteristic lines in the X-ray
diffraction spectra. The $\beta$-phase was also confirmed by the
resistivity of the films \cite{liu12}, which for 10~nm Ta thickness
lies at 200~$\mu\Omega\cdot$cm.

In order to compare ISHE and SMR on YIG$|$Pt and YIG$|$Ta bilayers, a
15~nm thick Pt and a 3~nm thick Ta layers were grown on the YIG1
sample. The conductivities of these metallic films are
$\sigma^\text{Pt}=2.45\cdot10^6~\Omega^{-1}\cdot$m$^{-1}$ (in
agreement with the values reported in
Refs.\cite{mosendz10a,castel12a}) and $\sigma^\text{Ta}=3.05
\cdot10^5~\Omega^{-1}\cdot$m$^{-1}$, respectively. These two samples
have been used to obtain the results presented in Figs.\ref{fig:2} and
\ref{fig:4}. The dependence on Pt thickness of both $V_\text{ISH}$
\cite{castel12a} and magnetoresistance \cite{huang12,vlietstra13} has
been studied earlier.  In this work, we have used the YIG2 sample to
study the dependence as a function of the Ta thickness, which was
varied from 1.5~nm to 15~nm (1.5, 2, 3, 5, 10 and 15~nm). The
conductivity of these Ta films increases from $0.8
\cdot10^5~\Omega^{-1}\cdot$m$^{-1}$ to $7.5
\cdot10^5~\Omega^{-1}\cdot$m$^{-1}$ with the film thickness. This
series of samples has been used to obtain the data of
Fig.\ref{fig:3}. Finally, Pt films with thicknesses 10~nm and 15~nm
were also grown on YIG2, for the sake of comparison with YIG1.

\subsection{Measurement setup}

A 500~$\mu$m wide, 2~$\mu$m thick Au transmission line cell and
electronics providing frequencies up to 20~GHz were used for microwave
measurements. The long axis of the sample was aligned perpendicularly
to the microwave line, thus parallel to the excitation field
$h_\text{rf}$ as indicated in the inset of
Fig.\ref{fig:2}. $V_\text{ISH}$ was measured by a lock-in technique
(with the microwave power turned on and off at a frequency of a few
kHz) with electrical connections through gold leads at equal distance
to the area of excitation. Magnetotransport measurements of the
YIG$|$NM slabs were performed using a 4-point configuration. The
samples were placed at the center of an electromagnet, which can be
rotated around its axis in order to obtain curves of magnetoresistance
vs. angle. The measurement cell was placed in a cryostat, with the
possibility to cool down to 77~K. All the measurements presented in
this paper were performed at room temperature, except for those
reported in Fig.\ref{fig:5}.

\section{Experimental results and analysis}

\subsection{Inverse spin Hall voltage: YIG$|$Pt vs. YIG$|$Ta}

\begin{figure}
  \includegraphics[width=\columnwidth]{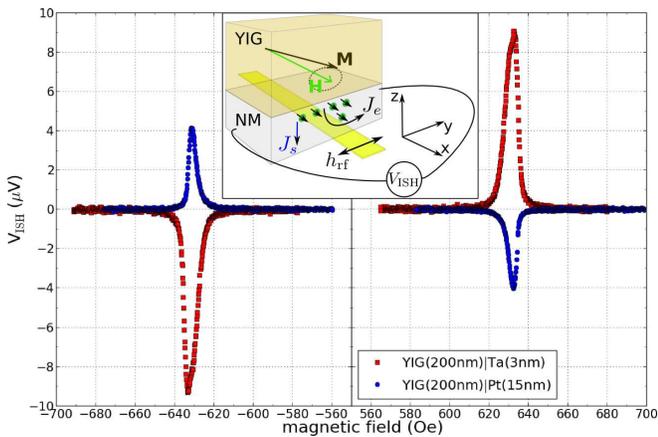}
  \caption{(Color online). Inverse spin Hall voltage measured at
    3.5~GHz for YIG$|$Ta and YIG$|$Pt. Inset: sketch of the
    experiment.}
  \label{fig:2}
\end{figure}

First, we compare in Fig.\ref{fig:2} the inverse spin Hall voltages
measured at 3.5~GHz ($P=+10$~dBm) in the YIG$|$Pt and YIG$|$Ta
bilayers. It shows that one can electrically detect the FMR of YIG in
these hybrid systems \cite{kajiwara10}. The spin current $J_s$ pumped
into the adjacent normal metal by the precessing magnetization in YIG
is converted into a charge current by ISHE,
\begin{equation}
  J_e=\frac{2e}{\hbar}\Theta_{SH}J_s \, ,
  \label{eq:jejs}
\end{equation}
where $e$ is the electron charge and $\hbar$ the reduced Planck
constant. This leads to a transverse voltage $V_\text{ISH}$ (across
the length of the YIG$|$NM slab), as sketched in the inset of
Fig.\ref{fig:2}. Moreover, $V_\text{ISH}$ must change sign upon
reversing the magnetization of YIG because of the concomitant reversal
of the spin pumped current $J_s$ (hence $J_e$). This is observed in
both the YIG$|$Pt and YIG$|$Ta systems, where $V_\text{ISH}$ is odd in
applied magnetic field, which shows that the voltage generated at
resonance is not due to a thermoelectrical effect.

The striking feature to be observed here is the opposite signs of
$V_\text{ISH}$ in these two samples. This remains true at all
microwave frequencies (from 2 to 8~GHz) and power levels (from $-$8 to
$+$10~dBm) which were measured, as well as for the different YIG$|$Pt
and YIG$|$Ta bilayers made from YIG1 and YIG2 samples. It thus
confirms that the spin Hall angles in Ta and Pt have opposite signs,
as predicted by ab initio calculations \cite{tanaka08} and inferred
from measurements where the spin current was generated by a metallic
ferromagnet \cite{morota11,liu12}. Moreover, from the electrical
circuit which was used in the measurements (anode of the voltmeter is
on the left in Fig.\ref{fig:2} inset), it can be found that
$\Theta_{SH}^\text{Pt}>0$ while $\Theta_{SH}^\text{Ta}<0$. The precise
estimation of the spin Hall angles in these two materials requires the
more analysis presented in the following sections. Still, it is
interesting to note that the 4~$\mu$V amplitude of $V_\text{ISH}$
measured in Fig.\ref{fig:2} on our 15~nm thick Pt is close to the one
reported in Ref.\cite{castel12a} (2 to 3~$\mu$V) with comparable
experimental conditions.

\subsection{Dependence of inverse spin Hall voltage on Ta thickness}

\begin{figure}
  \includegraphics[width=\columnwidth]{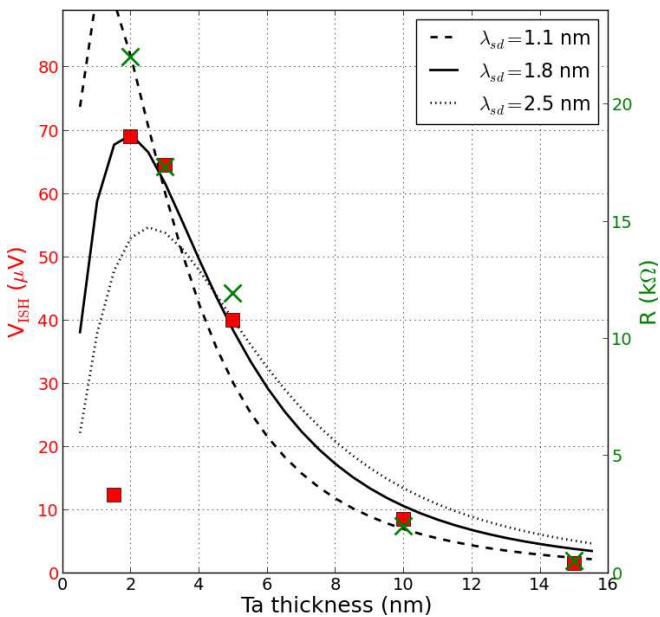}
  \caption{(Color online). Dependence of inverse spin Hall voltage on
    Ta thickness (red squares, left scale). The microwave frequency is
    3.8 GHz ($P=+10$~dBm). The lines are theoretical predictions
    \cite{castel12a} from Eq.\ref{eq:Vish} for different values of
    $\lambda_{sd}$, with the parameters $G_{\uparrow
      \downarrow}=4.3\cdot 10^{13}~\Omega^{-1}\cdot$m$^{-2}$ and
    $\Theta_{SH}=-0.02$. The resistance of the samples is also
    displayed (green crosses, right scale). The resistance of the 1.5
    nm thin Ta sample (95~k$\Omega$) is out-of-range.}
  \label{fig:3}
\end{figure}

In this work, we have measured the dependence of $V_\text{ISH}$ only
on Ta thickness. The study as a function of Pt thickness was already
reported in Ref.\cite{castel12a}, using a similar 200~nm thick YIG
film (fabricated in the same lab). In Fig.\ref{fig:3}, we have plotted
using red squares the dependence of $V_\text{ISH}$ on the Ta thickness
measured on the series of samples described above. Here,
$V_\text{ISH}$ is produced by the precession of magnetization in YIG,
resonantly excited at 3.8~GHz by the microwave field
($P=+10$~dBm). $V_\text{ISH}$ increases from less than 2~$\mu$V up to
$70~\mu$V as the Ta layer thickness is reduced from 15~nm to 2~nm, at
which the maximal voltage is measured. For the thinnest Ta layer
($t_\text{Ta}=1.5$~nm), $V_\text{ISH}$ drops to about 10~$\mu$V, a
value close to the one observed at $t_\text{Ta}=10$~nm. A similar
dependence of $V_\text{ISH}$ on Pt thickness was reported in
Ref.\cite{castel12a}, where a maximum of voltage was observed between
$t_\text{Pt}=1.5$~nm and $t_\text{Pt}=6$~nm.

The resistance measured across the length of the YIG$|$Ta slab is also
plotted with green crosses in Fig.\ref{fig:3} as a function of
$t_\text{Ta}$ (see right scale). It is interesting to note that both
$V_\text{ISH}$ and $R$ follow a similar dependence on the Ta
thickness, if one excludes the thinnest Ta layer, which might be
discontinuous or oxidised, and thus exhibits a very large resistance
($R=95$~k$\Omega$ is out-of-range of the graph).

To analyze the thickness dependence of the inverse spin Hall voltage,
we follow the approach derived in Ref.\cite{castel12a}. The spin
diffusion equation with the appropriate source term and boundary
conditions leads to the following expression:
\begin{eqnarray}
  V_\text{ISH} & = & \Theta_{SH} \frac{G_{\uparrow \downarrow}}{G_{\uparrow \downarrow}+\frac{\sigma}{\lambda_{sd}}\frac{1-\exp{(-2t_\text{NM}/\lambda_{sd}})}{1+\exp{(-2t_\text{NM}/\lambda_{sd})}}} \nonumber \\
  && \times \frac{h L P f \sin^2(\theta)}{2 e t_\text{NM}} \frac{(1-\exp{(-t_\text{NM}/\lambda_{sd}}))^2}{1+\exp{(-2t_\text{NM}/\lambda_{sd})}} \, ,
  \label{eq:Vish}
\end{eqnarray}
where $\sigma$ is the conductivity of the normal metal, $t_\text{NM}$
its thickness, $L$ the length of the YIG$|$NM slab excited at
frequency $f$ by the microwave field, $\theta$ the angle of precession
of YIG, and $P$ an ellipticity correction factor. The latter depends
on the excitation frequency \cite{mosendz10a} and in our case $P\simeq
1.25$.

From Eq.\ref{eq:Vish}, the \emph{amplitude} of $V_\text{ISH}$ depends
on the transport parameters $\lambda_{sd}$, $G_{\uparrow \downarrow}$
and $\Theta_{SH}$, as well as on the resonant precession angle
$\theta$. We do not have a direct measurement of $\theta$, but it can
be evaluated from the strength of the microwave field $h_\text{rf}$
and the measured linewidth $\Delta H$ \cite{gurevich96}. By performing
network analyzer measurements and considering the geometry of the
transmission line, we estimate the strength of the microwave field
$h_\text{rf}\simeq 0.2$~Oe for a $P=+10$~dBm output power from the
synthesizer. For the series of YIG$|$Ta samples, it yields a
precession angle $\theta\simeq3.3^\circ$ in YIG at
3.8~GHz. Nevertheless, the measurements presented in Fig.\ref{fig:3}
are not sufficient to extract independently $G_{\uparrow \downarrow}$
and $\Theta_{SH}$.

The \emph{thickness dependence} of $V_\text{ISH}$ primarily depends on
$\lambda_{sd}$, through the argument of the exponential functions in
Eq.\ref{eq:Vish}. The spin diffusion length can thus be adjusted to
fit the shape of $V_\text{ISH}$ vs. $t_\text{Ta}$ in
Fig.\ref{fig:3}. The series of lines in Fig.\ref{fig:3} displays the
result of calculations based on Eq.\ref{eq:Vish} for three different
values of $\lambda_{sd}$, using the thickness dependent conductivity
$\sigma^\text{Ta}$ measured experimentally. A very good overall
agreement to the data is found for a spin diffusion length
$\lambda_{sd}^\text{Ta}=1.8$~nm. We explain the discrepancy observed
at $t_\text{Ta}=1.5$~nm, at which the measured voltage is about five
times smaller than predicted, by the fact that the thinnest Ta layer
is discontinuous or oxidised, as already pointed out. We note that the
spin diffusion length extracted from the YIG$|$Ta data of
Fig.\ref{fig:3} is somewhat shorter than the 2.7~nm inferred from
nonlocal spin-valve measurements \cite{morota11}.

\subsection{Magnetoresistance: YIG$|$Pt vs. YIG$|$Ta}

\begin{figure}
  \includegraphics[width=\columnwidth]{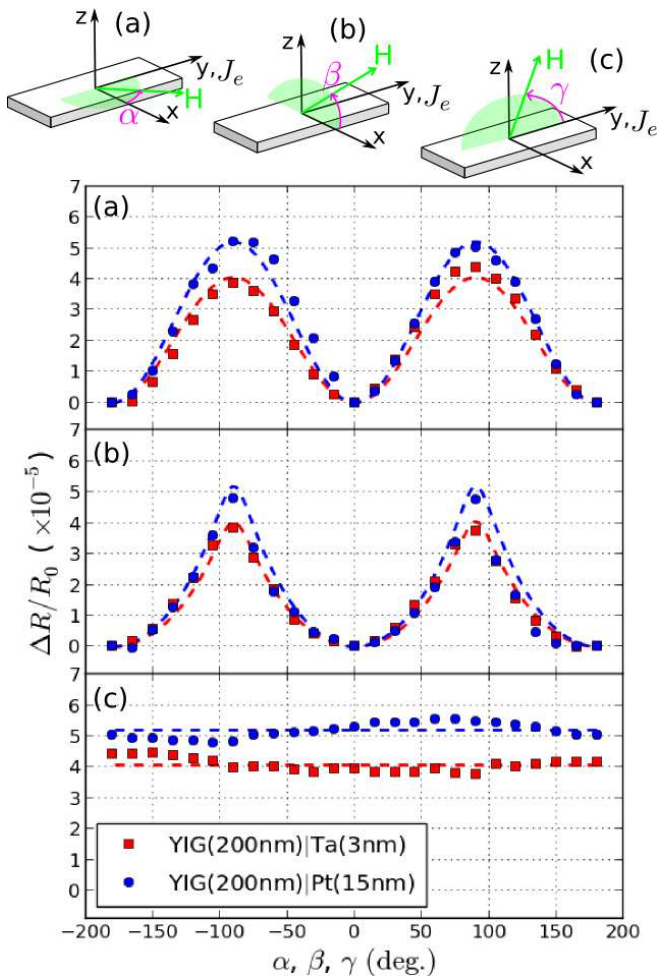}
  \caption{(Color online). (a$-$c) Magnetoresistance in YIG$|$Ta and
    YIG$|$Pt as a function of the angle of the applied field
    ($H=3$~kOe) sketched at the top (the samples are the same as the
    ones measured in Fig.\ref{fig:2}). Dashed lines are predictions
    from Eq.\ref{eq:SMR2} of the SMR theory \cite{nakayama12}.}
  \label{fig:4}
\end{figure}

We now turn to the measurements of dc magnetoresistance in our hybrid
YIG$|$NM bilayers. We have measured the variation of resistance in the
exact same samples as the ones studied by ISHE in Fig.\ref{fig:2},
YIG$|$Pt(15~nm) and YIG$|$Ta(3~nm), as a function of the angle of the
applied field with respect to the three main axes of the slabs. In
these experiments, the applied field was fixed to $H=3$~kOe
(sufficient to saturate the YIG), and a dc current of a few mA
together with a $6^{1/2}$ digits voltmeter were used to probe the
resistance of the NM layers in a 4-probe configuration. The results
obtained by rotating the magnetic field in the plane of the sample
(angle $\alpha$), from in-plane perpendicular to the charge current
$J_e$ to out-of-plane (angle $\beta$) and from in-plane parallel to
$J_e$ to out-of-plane (angle $\gamma$) are presented in
Figs.\ref{fig:4}a, \ref{fig:4}b, and \ref{fig:4}c, respectively (see
also associated sketches).

In both the YIG$|$Pt and YIG$|$Ta bilayers, we do observe some weak
magnetoresistance ($\Delta R_\text{max}/R_0$ of $5 \cdot 10^{-5}$ and
$4 \cdot 10^{-5}$, respectively), as it was first reported on the
YIG$|$Pt system \cite{huang12}. We checked that this weak variation
does not depend on the sign or strength of the probing current. In
contrast to the inverse spin Hall voltage measurements presented in
Fig.\ref{fig:2}, we also note that the sign (or symmetry) of the
effect is identical in YIG$|$Pt and YIG$|$Ta.

In order to interpret this magnetoresistance, it is important to
understand its dependence on all three different angles, $\alpha$,
$\beta$ and $\gamma$, shown in Fig.\ref{fig:4}. If one would just look
at the in-plane behavior (Fig.\ref{fig:4}a), one could conclude that
the NM resistance $R$ changes according to some anisotropic
magnetoresistance (AMR) effect, as if the NM would be magnetized at
the interface with YIG due to proximity effect \cite{huang12}. But
with AMR, $R$ depends on the angle between the charge current $J_e$
and the magnetization (applied field $H$). Hence, no change of $R$ is
expected with the angle $\beta$, whereas $R$ should vary with the
angle $\gamma$, which is exactly opposite to what is observed in
Figs.\ref{fig:4}b and \ref{fig:4}c, respectively. Therefore usual AMR
as the origin of the magnetoresistance in YIG$|$Pt and YIG$|$Ta
bilayers has to be excluded.

Instead, the spin Hall magnetoresistance (SMR) mechanism proposed in
Ref.\cite{nakayama12} is well supported by our magnetoresistance
data. In this scenario, the electrons carried by the charge current in
the NM layer are deflected by SHE in opposite directions depending on
their spin. Those whose spin is flipped by scattering at the interface
with the FM can oppose the initial current by ISHE and lead to an
increase of resistance. Therefore, the spin Hall magnetoresistance
depends on the relative angle between the magnetization $\bm{M}$ of
the FM and the accumulated spins $\bm{s}$ at the FM$|$NM interface:
\begin{equation}
  R=R_0+\Delta R_\text{max} \sin^2{(\bm{M},\bm{s})}\, .
  \label{eq:SMR2}
\end{equation}
The increase of resistance is maximal when $\bm{M}$ and $\bm{s}$ are
perpendicular, because the spin-flip-scattering governed by
$G_{\uparrow \downarrow}$ at the interface is the largest. In the
geometry depicted in Fig.\ref{fig:4}, the charge current is applied
along $y$, hence the spins accumulated at the YIG$|$NM interface due
to SHE are oriented along $x$. The dashed lines plotted in
Figs.\ref{fig:4}a$-$c are the prediction of the SMR theory. As can be
seen, Eq.\ref{eq:SMR2} explains well the presence (absence) of
resistance variation upon the applied field angles $\alpha$ and
$\beta$ ($\gamma$). Due to demagnetizing effects, the magnetization of
YIG is not always aligned with the applied field. This is the reason
why the measured curves in Figs.\ref{fig:4}a and \ref{fig:4}b have
different shapes, and a simple calculation \cite{gurevich96} of the
equilibrium position of $\bm{M}$ in combination with Eq.\ref{eq:SMR2}
reproduces them quite well.

The SMR ratio was also calculated in Ref.\cite{nakayama12}:
\begin{equation}
  \text{SMR}=\frac{\Delta R_\text{max}}{R_0} = \Theta_{SH}^2 \frac{\frac{2\lambda_{sd}^2}{\sigma t_\text{NM}}G_{\uparrow \downarrow}\tanh^2{\left(\frac{t_\text{NM}}{2\lambda_{sd}}\right)}}{1+\frac{2\lambda_{sd}}{\sigma}G_{\uparrow \downarrow}\coth{\left(\frac{t_\text{NM}}{\lambda_{sd}}\right)}} \, .
  \label{eq:SMR}
\end{equation}
As for the inverse spin Hall voltage $V_\text{ISH}$
(Eq.\ref{eq:Vish}), the SMR depends on all the transport parameters
$G_{\uparrow \downarrow}$, $\Theta_{SH}$ and $\lambda_{sd}$, which
therefore cannot be extracted individually from a single
measurement. In section \ref{sec:ratio}, we will take advantage of the
combined measurements of $V_\text{ISH}$ (Figs.\ref{fig:2} and
\ref{fig:3}) and SMR (Fig.\ref{fig:4}) to do so. For now, it is
interesting to point out that because both SHE and ISHE are at play in
spin Hall magnetoresistance, the SMR depends on the \emph{square} of
the spin Hall angle. This explains the positive SMR for both YIG$|$Pt
and YIG$|$Ta, even though the spin Hall angles of Pt and Ta are
opposite.

Finally, it would have been interesting to measure the dependence of
SMR on Ta thickness (the dependence on Pt thickness was studied in
Refs.\cite{huang12} and \cite{vlietstra13}). Unfortunately, it was
difficult to realize low noise 4-point contacts to investigate the
faint magnetoresistance on the series of Ta samples prepared to study
$V_\text{ISH}$ vs. $t_\text{Ta}$. From our attempts, we found that the
SMR of YIG$|$Ta(10~nm) is less than $2\cdot10^{-5}$. This is
consistent with the decrease predicted by Eq.\ref{eq:SMR} (assuming
$\lambda_{sd}^\text{Ta}=1.8$~nm) with respect to the $\text{SMR}\simeq
4\cdot10^{-5}$ measured for YIG$|$Ta(3~nm).

\section{Discussion}

\subsection{Transport parameters \label{sec:ratio}}

As already discussed, both $V_\text{ISH}$ and SMR depend on the set of
transport parameters ($G_{\uparrow \downarrow}$, $\Theta_{SH}$,
$\lambda_{sd}$). By studying $V_\text{ISH}$ as a function of the NM
thickness, the spin diffusion length can be determined, and we found
that in Ta, $\lambda_{sd}^\text{Ta}=1.8\pm0.7$~nm, see
Fig.\ref{fig:3}. We mention here that from a similar study on
YIG$|$Pt, $\lambda_{sd}^\text{Pt}=3.0\pm0.5$~nm could be inferred
\cite{castel12a}. This value lies in the range of spin diffusion
lengths reported on Pt, which span over almost an order of magnitude
\cite{liu11b}, from slightly more than 1~nm up to 10~nm.

There is a direct way to get the spin mixing conductance of a FM$|$NM
interface, by determining the increase of damping in the FM layer
associated to spin pumping in the adjacent NM layer
\cite{tserkovnyak05}. Due to its interfacial nature, this effect is
inversely proportional to the thickness of the FM and can be measured
only on ultra-thin films. This was recently achieved in nm-thick YIG
films grown by pulsed laser deposition \cite{heinrich11,burrowes12},
where spin mixing conductances $G_{\uparrow \downarrow}=(0.7-3.5)
\cdot 10^{14}~\Omega^{-1}\cdot$m$^{-2}$ have been reported for the
YIG$|$Au interface.

\begin{table*}
  \caption{Transport parameters obtained from the analysis of inverse spin Hall voltage (Figs.\ref{fig:2} and
    \ref{fig:3} $+$ Eq.\ref{eq:Vish}) and spin Hall magnetoresistance (Fig.\ref{fig:4} $+$ Eq.\ref{eq:SMR}) performed on YIG$|$Ta(1.5~nm$-$15~nm) and YIG$|$Pt(15~nm).}\label{tab:1}
\begin{ruledtabular}
\begin{tabular}{c c c}
  & YIG$|$Ta (1.5~nm $-$ 15~nm) & YIG$|$Pt (15~nm) \\
  \hline \\
  $\sigma$ ($10^6$ $\Omega^{-1}\cdot$m$^{-1}$) & $0.08 - 0.75$ & $2.45 \pm 0.10$ \\
  $\lambda_{sd}$ ($10^{-9}$~m) & $1.8\pm0.7$ & n/a [from 1.5 to 10]\cite{liu11b} \\
  $G_{\uparrow \downarrow}$ ($10^{13}~\Omega^{-1}\cdot$m$^{-2}$) & $4.3 \pm ^{11}_{2}$ & $6.2 \pm ^{14}_{4}$ \\
  $\Theta_{SH}$ & $-0.02 \pm ^{0.008}_{0.015}$ & $0.03 \pm ^{0.04}_{0.015}$ \\
\end{tabular}
\end{ruledtabular}
\end{table*}

Even for 200~nm thick YIG films as ours, it is possible to obtain the
full set of transport parameters thanks to our combined measurements
of $V_\text{ISH}$ and SMR on YIG$|$NM hybrid structures. In fact, from
Eqs.\ref{eq:Vish} and \ref{eq:SMR}, the ratio
$V_\text{ISH}^2/\text{SMR}$ does not depend on $\Theta_{SH}$, which
allows to determine $G_{\uparrow \downarrow}$. Then, the last unknown
$\Theta_{SH}$ can be found from the $V_\text{ISH}$ or SMR signal. This
is how we proceed to determine the transport parameters which are
collected in Table \ref{tab:1}. The drawback of this method is that it
critically relies on: i) $\lambda_{sd}$, which enters in the argument
of exponential functions in Eqs.\ref{eq:Vish} and \ref{eq:SMR}; and
ii) the angle of precession $\theta$ in the inverse spin Hall
experiment, since $V_\text{ISH}^2/\text{SMR}\propto \theta^4$. Our
estimation of $\theta$ being within $\pm25\%$, the value extracted for
$G_{\uparrow \downarrow}$ from the ratio $V_\text{ISH}^2/\text{SMR}$
can vary by a factor up to 8 due to this uncertainty. The spin Hall
angle $\Theta_{SH}$ is less sensitive to other parameters, still it
can vary by a factor up to 3. This explains the rather large error
bars in Table \ref{tab:1}. In this study, we did not determine the
spin diffusion length in Pt, hence we used the range of values
reported in the literature \cite{liu11b}.

The spin mixing conductances determined from our combined
$V_\text{ISH}$ and SMR measurements on YIG$|$Ta and YIG$|$Pt bilayers
lie in the same window as the ones determined from interfacial
increase of damping in YIG$|$Au \cite{heinrich11}, from inverse spin
Hall voltage in BiY$_2$Fe$_5$O$_{12}|$Au and Pt \cite{takahashi12},
and from first-principles calculations in YIG$|$Ag \cite{jia11}.  We
would like to point out that despite the large uncertainty,
$G_{\uparrow \downarrow}$ for YIG$|$Ta is likely less than for
YIG$|$Pt. We note that the smaller damping measured in CoFeB$|$Ta
compared to CoFeB$|$Pt was tentatively attributed to a smaller spin
mixing conductance \cite{liu12}.

The spin Hall angles that we report for Pt and Ta are both of a few
percents. In particular, $\Theta_{SH}^\text{Ta}\simeq-0.02$ lies in
between the values determined from nonlocal spin-valve measurements
($\simeq-0.004$) \cite{morota11} and from spin-torque switching using
the SHE ($\simeq-0.12$) \cite{liu12}.

The main conclusion which arises from the summary presented in Table
\ref{tab:1} is that the sets of transport parameters determined for
the hybrid YIG$|$Ta and YIG$|$Pt systems are quite similar. Apart from
the opposite sign of $\Theta_{SH}$ in Ta and Pt, the main difference
concerns the conductivity: $\sigma^{\beta-\text{Ta}}$ is roughly one
order of magnitude smaller than $\sigma^\text{Pt}$. This explains the
large inverse spin Hall voltages that can be detected in our YIG$|$Ta
bilayers (up to 70~$\mu$V at $P=+10$~dBm), since from Eq.\ref{eq:Vish}
$V_\text{ISH} \propto 1/\sigma$, which could be a useful feature of
the Ta layer.

\subsection{Influence of a dc current on FMR linewidth}

Onsager reciprocal relations imply that if there is an ISHE voltage
produced by the precession of YIG, there must also be a transfer of
spin angular momentum from the NM conduction electrons to the
magnetization of YIG, through the finite spin mixing conductance at
the YIG$|$NM interface \cite{jia11}. Therefore, one would expect to be
able to control the relaxation of the insulating YIG by injecting a dc
current in an adjacent strong spin-orbit metal, as it was shown on
YIG$|$Pt in the pioneering work of Kajiwara et
al. \cite{kajiwara10}. Although this direct effect is well established
when the ferromagnetic layer is ultra-thin and metallic
\cite{ando08a,demidov11d,demidov12,liu12b}, only a few works report on
conclusive effects on micron-thick YIG
\cite{kajiwara10,wang11,padron-hernandez11} or provide a theoretical
interpretation to the phenomenon \cite{xiao12}.

The 200~nm thick YIG films that have been grown for this study are
about 6 times thinner than the one used in Ref.\cite{kajiwara10}, with
an intrinsic relaxation close to bulk YIG. Because the spin transfer
torque is an interfacial effect and sizable spin mixing conductances
have been measured in our YIG$|$Ta and YIG$|$Pt bilayers, our samples
must be good candidates to observe the direct effect of a dc current
on the relaxation of YIG. Due to their large resistance, $\beta$-Ta
films are not convenient to pass the large current densities required
to observe such an effect (large Joule heating). Therefore, we have
conducted these experiments only on the YIG$|$Pt films prepared in
this work.

\begin{figure}
  \includegraphics[width=\columnwidth]{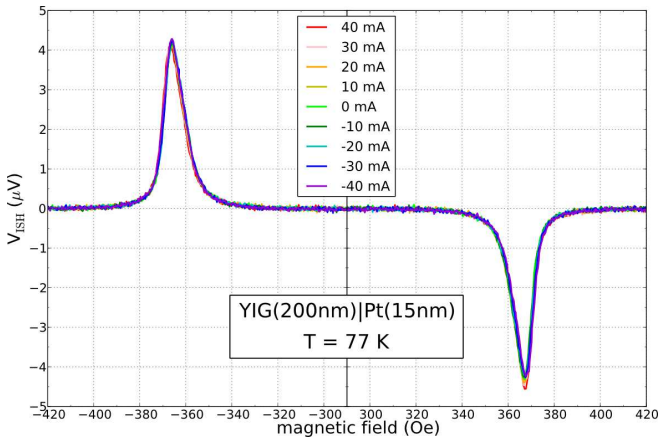}
  \caption{(Color online). Inverse spin Hall voltage measured at
    2.95~GHz ($P=+10$~dBm) for YIG$|$Pt as a function of the dc
    current flowing in the Pt layer. A small current dependent offset
    ($<0.2~\mu$V) has been subtracted to the data.}
  \label{fig:5}
\end{figure}

The inverse spin Hall voltage measurements presented in
Fig.\ref{fig:2} have therefore been repeated in presence of a dc
current flowing through the Pt layer. This type of experiment, where a
ferromagnetic layer is excited by a small amplitude signal and a spin
polarized current can influence the linewidth of the resonance, has
already been reported on spin-valve spin-torque oscillators
\cite{sankey06,chen08} and NiFe$|$Pt bilayers \cite{liu11}. The
results obtained on our YIG(200~nm)$|$Pt(15~nm) at 77~K when the dc
current is varied from $-40$ to $+40$~mA are displayed in
Fig.\ref{fig:5}.

Let us now comment on these experiments. We first emphasize that the
current injected in Pt is truly dc (not pulsed). A sizable Joule
heating is thus induced, as reflected by the increase of Pt
resistance. As a consequence, the main effect of dc current injection
at room temperature is the displacement of the resonance towards
larger field, due to the decrease of the YIG saturation magnetization
$M_s$. To avoid this trivial effect, we have performed these
experiments directly in liquid nitrogen. In that case, the increase of
Pt resistance is very limited ($+0.2$\% at $\pm40$~mA). We note that
when cooled from 300~K down to 77~K, the peak of the inverse spin Hall
voltage measured in the YIG$|$Pt bilayer is displaced towards lower
field due to the increase of $M_s$ of YIG (from 140~emu/cm$^3$ up to
200~emu/cm$^3$), and its amplitude slightly decreases.

The main conclusion that can be drawn from Fig.\ref{fig:5} is that
there is basically no effect of the dc current injected in Pt on the
YIG resonance. We stress that the maximal current density reached in
Pt in these experiments is $J_e=2.4\cdot10^9$~A.m$^{-2}$,
\textit{i.e.}, twice larger than the one at which YIG magnetization
oscillations were reported in Ref.\cite{kajiwara10}. In our
experiments, we are not looking for auto-oscillations of YIG, which
requires that the damping is fully compensated by spin transfer
torque, but only for some variation of the linewidth. The fact that we
do not see any change in the shape of the resonant peak of our 200~nm
thin YIG film is thus in contradiction with the observation of bulk
auto-oscillations in thicker films \cite{kajiwara10}.

We have also performed similar experiments on the other
YIG(200~nm)$|$Pt samples which were prepared using the two different
YIG films grown for this study. Although the current density was
increased up to $6\cdot10^9$~A.m$^{-2}$, we were never able to detect
any sizable variation of the linewidth of YIG. Instead, we have
measured that the dc current can affect the inverse spin Hall voltage
in different ways. First, when a charge current is injected into Pt, a
non-zero offset of the lock-in signal can be detected (it was
subtracted in Fig.\ref{fig:5}). This is due to the increase of Pt
resistance induced by the microwave power, as it was verified by
monitoring this offset while varying the modulation frequency of the
microwave. Secondly, the amplitude of the $V_\text{ISH}$ peaks can be
affected by the dc current (but again, \emph{not} the linewidth). This
effect can at first be confused with some influence on the relaxation
of YIG, because it displays the appropriate symmetries vs. field and
current. But instead, we have found that this is a bolometric
effect\cite{gui07}: when the YIG is excited at resonance, it heats up,
thereby heating the adjacent Pt whose resistance gets slightly
larger. Hence an additional voltage to $V_\text{ISH}$ is picked up on
the lock-in due to the non-zero dc current flowing in Pt. Therefore,
one should be very careful in interpreting changes in inverse spin
Hall voltage as the indication of damping variation in YIG. Finally,
we observed that at very large current density, the resonance peak
slightly shifts towards larger field due to Joule heating, even at
77~K.

\section{Conclusion}

In this paper, we have presented and analyzed a comparative set of
data of inverse spin Hall voltage $V_\text{ISH}$ and magnetoresistance
obtained on YIG$|$Pt and YIG$|$Ta bilayers. We have detected the
voltages generated by spin pumping at the YIG$|$Pt interface (already
well established \cite{kajiwara10}) and at the YIG$|$Ta interface (for
the first time). Their opposite signs are assigned to the opposite
spin Hall angles in Pt and Ta \cite{tanaka08}.  From the thickness
dependence of $V_\text{ISH}$, we have been able to obtain the spin
diffusion length in Ta, $\lambda_{sd}^\text{Ta}=1.8\pm0.7$~nm, in
reasonable agreement with the value extracted from non-local spin
valve measurements \cite{morota11}. From symmetry arguments, we have
shown that the weak magnetoresistance measured on our hybrid YIG$|$NM
layers cannot be attributed to usual AMR, but is instead well
understood in the framework of the recently introduced spin Hall
magnetoresistance (SMR) \cite{nakayama12}. By taking advantage of the
combined measurements of $V_\text{ISH}$ and SMR performed on the same
samples, we have been able to extract the spin Hall angles in Pt and
Ta, as well as the spin mixing conductances at the YIG$|$Pt and
YIG$|$Ta interfaces.

These transport parameters have all been found to be of the same order
of magnitude as those already measured \cite{liu12,heinrich11} or
predicted \cite{jia11}. We believe that at least part of the
discrepancies between the parameters evaluated in different works
\cite{liu11b} depend on the details of the YIG$|$NM interface
\cite{burrowes12} and on the quality of the NM
\cite{mosendz10a,kondou12,morota11}.

Finally, we could not detect any change of linewidth in our YIG$|$Pt
samples by passing large current densities through the Pt layer. One
might argue that our high quality 200~nm YIG thin films are still too
thick to observe any appreciable effect of spin transfer torque, which
is an interfacial mechanism, or that the spin-waves which can
auto-oscillate under the action of spin transfer at the interface with
Pt are different from the uniform mode that we excite with the
microwave field in our experiments \cite{kajiwara10,xiao12}. If one
would estimate the threshold current required to fully compensate the
damping of all the magnetic moments contained in our YIG films
\cite{liu12,xiao12}, $J_\text{th} \simeq 2e\alpha\omega
M_st_\text{YIG}/(\Theta_{SH} \gamma \hbar)$, one would get current
densities of about $10^{11}$~A.m$^{-2}$. This is 20 times larger than
the largest current density which we have tried. Thus the lack of a
visible effect in our Fig.\ref{fig:5} is not a real surprise in
itself, but it is inconsistent with the results reported in
Ref.\cite{kajiwara10}. Future experiments on ultra-thin YIG$|$NM
hybrid films, in which the spin mixing conductance can be directly
determined from the interfacial increase of damping \cite{heinrich11},
might give a definite answer to this point.


\begin{acknowledgments}
  This research was supported by the French ANR Grant Trinidad (ASTRID
  2012 program).
\end{acknowledgments}


%

\end{document}